\documentclass[a4paper,11pt]{article}
\usepackage[dvips,a4paper,text={16cm,24cm}]{geometry}
\usepackage[latin1]{inputenc}
\usepackage{amssymb}
\usepackage{amsmath}
\usepackage{graphicx}
\usepackage{pstricks}
\usepackage{epic}
\numberwithin{equation}{section}

\newcommand{\be}{\begin{equation}}
\newcommand{\ee}{\end{equation}}
\newcommand{\bea}{\begin{eqnarray}}
\newcommand{\eea}{\end{eqnarray}}

\newcommand{\G}{\left}
\newcommand{\D}{\right}

\newcommand{\bpsi}{\overline{\psi}}

\def\a{\alpha}
\def\b{\beta}
\def\g{\gamma}
\def\d{\delta}

\def\m{\mu}

\def\nn{\nnnumber}

\def\nn{\nonumber}

\begin{document}

\begin{flushright}
ULB-TH/05-29\\
December 2005\\
\vspace*{1cm}
\end{flushright}

\begin{center}
\begin{Large}
\textbf{Extended $E_8$ Invariance of $11$-Dimensional Supergravity}
\end{Large}

\vspace{7mm} {\bf Sophie de Buyl\footnote{Aspirant du Fonds
National de la Recherche Scientifique, Belgique}, Marc
Henneaux\footnote{Also at CECS, Valdivia, Chile}, Louis Paulot}

\vspace{5mm}
Physique th\'eorique et math\'ematique, Universit\'e libre de Bruxelles\\
and\\
International Solvay Institutes\\
Campus Plaine C.P.~231, B--1050 Bruxelles, Belgium \vspace{3mm}

\end{center}

\vspace{3mm} \hrule
\begin{abstract}

The hyperbolic Kac-Moody algebra $E_{10}$ has repeatedly been
suggested to play a crucial role in the symmetry structure of
$M$-theory. Recently, following the analysis of the asymptotic
behaviour of the supergravity fields near a cosmological
singularity, this question has received a new impulse.  It has been
argued that one way to exhibit the symmetry was to rewrite the
supergravity equations as the equations of motion of the non-linear
sigma model $E_{10}/K(E_{10})$.   This attempt, in line with the
established result that the scalar fields which appear in the
toroidal compactification down to three spacetime dimensions form
the coset $E_8/SO({16})$, was verified for the first bosonic levels
in a level expansion of the theory. We show that the same features
remain valid when one includes the gravitino field.

\end{abstract}
\hrule

\vspace{3mm}

\section{Introduction}

The hyperbolic Kac-Moody algebra $E_{10}$, whose Dynkin diagram is
given in Fig.1,  has repeatedly been argued to play a crucial role
in the symmetry structure of $M$-theory
\cite{JuliaInf,Nicolai,Mizo}.
\newline
\begin{center}
\ \scalebox{.8}{
\begin{picture}(180,60)
%nom des racines
\put(-55,-5){$\alpha_{9}$} \put(-15,-5){$\alpha_8$}
\put(25,-5){$\alpha_7$} \put(65,-5){$\alpha_6$}
\put(105,-5){$\alpha_5$} \put(145,-5){$\alpha_4$}
\put(185,-5){$\alpha_3$} \put(225,-5){$\alpha_2$}
\put(265,-5){$\alpha_1$} \put(200,45){$\alpha_0$}
%9 vertex + lignes simples
\thicklines \multiput(-50,10)(40,0){9}{\circle{10}}
\multiput(-45,10)(40,0){8}{\line(1,0){30}}
%1 vertex du dessus
\put(190,50){\circle{10}} \put(190,15){\line(0,1){30}}
\end{picture}
}
\end{center}

\centerline{
{\footnotesize FIG. 1.}
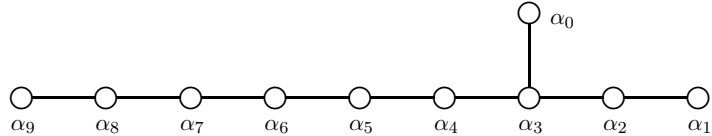
{\small \emph{The Dynkin diagram of $E_{10}$} }}
$ \ $ 

This infinite-dimensional algebra has a complicated structure that
has not been deciphered yet. In order to analyse further its root
pattern, it was found convenient in \cite{DHN} to introduce a
``level'' for any root $\a$, defined as the number of times the
simple root $\a_0$ occurs in the decomposition of $\a$.

The roots $\a_1$ through $\a_9$ define a subalgebra $sl(10)$.
Reflections in these roots define the finite Weyl group $W_{A_9}$
($\simeq S_{10}$) of $A_9$, which acts naturally on the roots of
$E_{10}$. If we express the roots of $E_{10}$ in terms of the
spatial scale factors $\b^i$ appearing naturally in cosmology
\cite{DH1}, the action of $W_{A_9}$ is simply to permute the $\b$'s.
The level is invariant under $W_{A_9}$.  Consider the set $R_{E_8}$
of roots of the $E_8$ subalgebra associated with the simple roots
$\a_0$ through $\a_7$. By acting with $W_{A_9}$ on $R_{E_8}$, one
generates a larger set $\tilde{R}_{E_8}$ of roots. This set will be
called the extended set of roots of $E_8$. By construction, the
roots in $\tilde{R}_{E_8}$ are all real and have length squared
equal to $2$. There is an interesting description of the roots in
$\tilde{R}_{E_8}$ in terms of the level.  One can easily verify that
all the roots at level $0$, $\pm 1$ and $\pm 2$, as well as all the
{\it real} roots at level $\pm 3$ exhaust $\tilde{R}_{E_8}$.  This
includes, in particular, all the roots with $\vert$height$\vert$
$<30$.

Recently, following the analysis {\it \`a la} BKL \cite{BKL,DHN2} of
the asymptotic behaviour of the supergravity fields near a
cosmological singularity, the question of the hidden symmetries of
eleven-dimensional supergravity has received a new impulse
\cite{DH1}. It has been argued that one way to exhibit the symmetry
was to rewrite the supergravity equations as the equations of motion
of the non-linear sigma model $E_{10}/K(E_{10})$ \cite{DHN}.

The first attempt for rewriting the equations of motion of
eleven-dimensional supergravity as non-linear sigma model equations
of motion -- in line with the established result that the scalar
fields which appear in the toroidal compactification down to three
spacetime dimensions form the coset $E_8/SO({16})$ \cite{CJ0} -- is
due to \cite{West}. In that approach, it is the larger
infinite-dimensional algebra $E_{11}$ which is priviledged. Various
evidence supporting $E_{11}$ was provided in \cite{West,EHTW}. Here,
we shall stick to (the subalgebra) $E_{10}$, for which the dynamical
formulation is clearer.

The idea of rewriting the equations of motion of eleven-dimensional
supergravity as equations of motion of $E_{10}/K(E_{10})$  was
verified in \cite{DHN} for the first bosonic levels in a level
expansion of the theory. More precisely, it was verified that in the
coset model $E_{10}/K(E_{10})$, the fields corresponding to the
Cartan subalgebra and to the positive roots $\in \tilde{R}_{E_8}$
have an interpretation in terms of the (bosonic) supergravity fields
(``dictionary" of \cite{DHN}). {}Furthermore, there is a perfect
match of the supergravity equations of motion and the coset model
equations of motion for the fields corresponding to these real
roots.  This extended $E_8$-invariance, which combines the known
$E_8$-invariance and the manifest $sl_{10}$-invariance, is a first
necessary step in exhibiting the full $E_{10}$ symmetry. Further
indication on the meaning of the fields associated with the higher
roots in terms of gradient expansions, using partly information from
$E_9$, was also given in \cite{DHN}.

The purpose of this paper is to explicitly verify the extended
$E_8$-invariance of the fermionic sector of $11$-dimensional
supergravity.  This amounts to showing that up to the requested
level, the fermionic part of the supergravity Lagrangian, which is
first order in the derivatives, can be written as \be i \Psi^T M D_t
\Psi \label{covariant}\ee where (i) $\Psi$ is an infinite object
that combines the spatial components of the gravitino field $\psi_a$
and its successive gradients \be \Psi = (\psi_a, \cdots) \ee in such
a way that $\Psi$ transforms in the representation of $K(E_{10})$
that reduces to the spin 3/2 representation of $SO(10)$; (ii) $M$ is
a $K(E_{10})$-invariant (infinite) matrix; and (iii) $D_t$ is the
$K(E_{10})$ covariant derivative. [We work in the gauge $\psi'_0=0$,
where $\psi'_0$ is the redefined temporal component of the gravitino
field familiar from dimensional reduction \cite{CJ0}, \be \psi'_0 =
\psi_0 - \g_0 \g^a \psi_a ,\label{redef}\ee so that the temporal
component of $\psi_\mu$ no longer appears.]

In fact, for the roots considered here, one can truncate $\Psi$ to
the undifferentiated components $\psi_a$.  The next components --
and the precise dictionary yielding their relationship with the
gravitino field gradients -- are not needed. In view of the fact
that the undifferentiated components of the gravitino field form a
representation of the maximal compact subgroup $SO(16)$ of $E_8$
in the reduction to three dimensions, without the need to
introduce gradients or duals, this result is not unexpected.

The crux of the computation consists in constructing the
representation of $K(E_{10})$ up to the required level. This is done
in the next section, where we compare and contrast the spin 1/2 and
spin 3/2 representations of $K(E_{10})$.  The technically simpler
case of the spin 1/2 representation was investigated in
\cite{deBuylHP}, where it was shown that the Dirac Lagrangian was
compatible with extended $E_8$ invariance provided one introduces an
appropriate Pauli coupling with the $3$-form.  Our work overlaps the
work \cite{West2} on the fermionic representations of $K(E_{11})$ as
well as the analyses of \cite{Keur3} on the maximal compact
subgroups of $E_{n,n}$ and of \cite{NicCompact} on $K(E_{9})$.

We then investigate the conjectured infinite-dimensional symmetry
$E_{10}$ of the Lagrangian of \cite{Cremmer:1978km}.   We find that
the fermionic part also takes the form dictated by extended
$E_8$-invariance, with the correct covariant derivatives appearing
up to the appropriate level. As observed by previous authors and in
particular in \cite{deBuylHP}, there is an interesting interplay
between supersymmetry and the hidden symmetries.

\section{`Spin 3/2' Representation of $K(E_{10})$}

\subsection{Level 0}

To construct the `spin 3/2' representation of $K(E_{10})$, we have
to extend the level 0 part which is the usual $SO(10)$ `spin 3/2'
parametrized by a set of 10 spinors $\chi_m$, where $m=1\ldots 10$
is a space index. [The level is not a grading for $KE_{10}$ but a
filtration, defined modulo lower order terms.]  The $so(10)$
generators $k^{ij}$ act on $\chi_m$ as \be k^{ij}.\chi_m =
\frac{1}{2} \gamma^{ij} \chi_m + \delta_{m}^{\; \;i} \chi^j -
\delta_{m}^{\; \; j} \chi^i \rlap{\ .} \label{k-ab}\ee The aim is
to rewrite the Rarita-Schwinger term with all couplings of the
fermionic field, up to higher order fermionic terms, into the form
(\ref{covariant}).

\subsection{Level 1}

Beyond $SO(10)$, the first level couples to $F_{0abc}$. To
reproduce the supergravity Lagrangian, the level 1 generators must
contain products of $\gamma$ matrices where the number of matrices
is odd and at most five. Indeed, the matrix $M$ in
(\ref{covariant}) is proportional to the antisymmetric product
$\g^{ab}$ (as one sees by expanding the supergravity Lagrangian
${\cal L} \sim i \psi^T_a \g^{ab} \dot{\psi}_b + \cdots$), while
$F_{0abc}$ is coupled to fermions, in the supergravity Lagrangian,
through terms $\psi_m^T \g^{mabcn}\psi_n$ and $\psi_a^T \g^{bc}
\g^n\psi_n$. In addition, the generators must be covariant with
respect to $SO(10)$. This gives the general form \be
k^{abc}.\chi_m = A \gamma_m^{\; \;nabc} \chi_n + 3 B \delta_m^{[a}
\gamma^{bc]n} \chi_n + 3C \gamma_m^{\;\;[ab}\chi^{c]} + 6D
\delta_m^{[a} \gamma^b \chi^{c]} + E \gamma^{abc} \chi_m
\label{gen-kabc} \ee where $A,B,C,D,E$ are constants to be fixed.
In fact, it is well known that such generators do appear in the
dimensional reduction of supergravity. If $d$ dimensions are
reduced, the generators  mix only $\chi_m$ with $1 \leq m \leq d$.
Therefore we set the terms involving summation on $n$ on the right
hand side of (\ref{gen-kabc}) equal to zero: $A=B=0$.

To fix the coefficients $C$, $D$ and $E$, we must check the
commutations relations. Commutation with level 0 generators is
automatic, as (\ref{gen-kabc}) is covariant with respect to
$SO(10)$. What is non-trivial is commutation of the generators at
level 1 with themselves. The generators $K^{abc}$ of $K(E_{10})$
at level 1 fulfill \be [K^{abc}, K^{def}] = K^{abcdef} - 18
\d^{ad} \d^{be} K^{cf} \ee (with antisymmetrization in $(a,b,c)$
and $(b,c,d)$ in $\d_{ad} \d_{be} K_{cf}$) as it follows from the
$E_{10}$ commutation relations. In order to have a representation
of $K(E_{10})$, the $k^{abc}$ must obey the same algebra, \be
[k^{abc}, k^{def}] = k^{abcdef} - \d^{ad} \d^{be} k^{cf}.
\label{level1algebra}\ee When all the indices are distinct,
(\ref{level1algebra}) defines the generators at level 2. One gets
non trivial constraints when two or more indices are equal.
Namely, there are two relations which must be imposed: \bea
\G[ k^{abc} , k^{abd} \D] &=& -k^{cd} \label{rel1}\\
\G[ k^{abc} , k^{ade} \D] &=& 0 \label{rel2}\eea where different
indices are supposed to be distinct. In fact, as we shall discuss in
the sequel, all other commutation relations which have to be checked
for higher levels can be derived from this ones using the Jacobi
identity. One can verify (\ref{rel1}) and (\ref{rel2}) directly or
using \emph{FORM} \cite{form}.  One finds that these two relations
are satisfied if and only if \be C=- \frac{1}{3} \epsilon , \; \; \;
\; D=\frac{2}{3} \epsilon, \; \; \; \;  E=\frac{1}{2} \epsilon \ee
with $\epsilon = \pm 1$. In fact one can change the sign of
$\epsilon$ by reversing the signs of all the generators at the odd
levels. This does not change the algebra. We shall use this freedom
to set $\epsilon =  1$ in order to match the conventions for the
supergravity Lagrangian. Putting everything together, the level 1
generator is \be k^{abc}.\chi_m =
 \frac{1}{2}  \gamma^{abc} \chi_m -
\gamma^{m[ab}\chi^{c]} +4  \delta_m^{[a} \gamma^b \chi^{c]}
\rlap{\ .} \label{k-abc} \ee

\subsection{Level 2}

The expression just obtained for the level 1 generators can be used
to compute the level 2 generator \be k^{abcdef} = \G[ k^{abc} ,
k^{def} \D] \ee which is totally antisymmetric in its indices, as it
can be shown using the Jacobi identity. Explicitly, Eq.(\ref{k-abc})
gives \be k^{abcdef}.\chi_m = \frac{1}{2} \gamma^{abcdef} \chi_m +4
\gamma_m^{\; \;[abcde} \chi^{f]} -10 \delta_m^{[a} \gamma^{bcde}
\chi^{f]} \rlap{\ .} \label{k-abcdef}\ee

\subsection{Level 3}

We now turn to level 3.  There are two types of roots. Real roots
have generators \be k^{a;abcdefgh} =  \G[ k^{abc} , k^{adefgh} \D]
\ee (without summation on $a$ and all other indices distinct).
They are easily computed to act as \be k^{a;abcdefgh}.\chi_m =
\frac{1}{2}  \gamma^{bcdefgh} \chi_m + 2 \gamma_m^{\; \; abcdefgh}
\chi^a +16  \delta_m^a \gamma^{[abcdefg}\chi^{h]} -7 \gamma_m^{\;
\; [bcdefg}\chi^{h]} \rlap{\ .} \label{real}\ee In addition, there
are generators $k^{a;bcdefghi}$ with all indices distinct,
corresponding to null roots. {}From \be \G[ k^{abc} , k^{defghi}
\D] = 3 k^{[a;bc]defghi} \ee (with all indices distinct) one finds
\be k^{a;bcdefghi}.\chi_m = - 2  \G( \gamma_m^{\;
\;[abcdefgh}\chi^{i]} - \gamma_m^{\; \; bcdefghi}\chi^a \D) - 16
\G( \delta_m^{[a} \gamma^{bcdefgh} \chi^{i]} - \delta_m^a
\gamma^{[bcdefgh}\chi^{i]} \D) \rlap{\ .} \label{null}\ee
Combining these results, one finds that the level 3 generators can
be written as \be
\begin{split}
k^{a;bcdefghi}.\chi_m = -2 & \G( \gamma_m^{\; \;
[abcdefgh}\chi^{i]} - \gamma_m^{\; \; bcdefghi}\chi^a \D) - 16 \G(
\delta_m^{[a} \gamma^{bcdefgh} \chi^{i]} - \delta_m^a
\gamma^{[bcdefgh}\chi^{i]} \D)\\& + 4  \delta^{a[b}
\gamma^{cdefghi]} \chi_m - 56
\gamma^{m[bcdefg}\delta^{\hat{a}h}\chi^{i]} \rlap{\ .}
\end{split} \label{level3}
\ee (where the hat over $a$ means that it is not involved in the
antisymmetrization).  Note that if one multiplies the generator
(\ref{level3}) by a parameter $\m_{a;bcdefghi}$ with the symmetries
of the level 3 Young tableau (in particular, $\m_{[a;bcdefghi]} =
0$), the first terms in the two parentheses disappear.
{}Furthermore, the totally antisymmetric part of the full level 3
generator vanishes. The condition $\m_{[a;bcdefghi]}=0$ on $\m_{a;bcdefghi}$ is equivalent
to the tracelessness
of its dual.  
\subsection{Compatibility checks}
Having defined the generators of the `spin 3/2' representation up
to level 3, we must now check that they fulfill all the necessary
compatibility conditions expressing that they represent the
$K(E_{10})$ algebra up to that level (encompassing the
compatibility conditions (\ref{rel1}) and (\ref{rel2}) found
above). This is actually a consequence of the Jacobi identity and
of the known $SO(16)$ invariance in 3 dimensions, as well as of
the manifest spatial $SL(10)$ covariance that makes all spatial
directions equivalent.

Consider for instance the commutators of level 1 generators with
level 2 generators.  The $K(E_{10})$ algebra is
 \be [K^{abc}, K^{defghi}] = 3 K^{[a;bc]defghi} - 5! \d^{ad}
\d^{be} \d^{cf} K^{ghi} \label{1-2}\ee Thus, one must have \be
[k^{abc}, k^{defghi}] = 3 k^{[a;bc]defghi} - 5! \d^{ad} \d^{be}
\d^{cf} k^{ghi} \label{1-2bis}\ee These relations are constraints
on $k^{abc}$ and $k^{defghi}$ when the level 3 generators are
absent, which occurs when (at least) two pairs of indices are
equal.  But in that case, there are only (at most) 7 distinct
values taken by the indices and the relations are then part of the
known $SO(16)$ invariance emerging in 3 dimensions. In fact, the
relations (\ref{1-2bis}) are known to hold when the indices take
at most 8 distinct values, which allows $k^{a;acdefghi}$ with a
pair of repeated indices. These 8 values can be thought of as
parameterizing the 8 transverse dimensions of the dimensional
reduction. Note that since the index $m$ in (\ref{k-abc}) can be
distinct from the 8 ``transverse" indices, we have both the `spin
1/2' and the `spin 3/2' (i.e., the vector and the spinor)
representations of $SO(16)$, showing the relevance of the analysis
of \cite{deBuylHP} in the present context.

Similarly, the commutation of two level 2 generators read \be
[K^{abcdef},K^{ghijkl}] = - 6\cdot 6! \d^{ag} \d^{bh} \d^{ci}
\d^{dj} \d^{ek} K^{fl} + \hbox{ ``more"} \ee where ``more" denotes
level 4 generators.  Thus, one must have \be
[k^{abcdef},k^{ghijkl}] = - 6\cdot 6! \d^{ag} \d^{bh} \d^{ci}
\d^{dj} \d^{ek} k^{fl} + \hbox{ ``more"} \label{2-2}\ee  These
relations are constraints when the level 4 generators are
absent\footnote{When the level 4 generators are present, the
relations (\ref{2-2}) are consequences of the definition of the
level 4 generators -- usually defined through commutation of level
1 with level 3 --, as a result of the Jacobi identity.}. Now, the
level 4 generators are in the representation $(001000001)$
characterized by a Young tableau with one 9-box column and one
3-box column, and in the representation $(200000000)$
characterized by a Young tableau with one 10-box column and two
1-box columns \cite{NicFisch,Fisch}.  To get rid of these level 4
representations, one must again assume that the indices take at
most 8 distinct values to have sufficiently many repetitions.  [If
one allows 9 distinct values, one can fill the tableau
$(001000001)$ non trivially.] But then, $SO(16)$  ``takes over"
and guarantees that the constraints are fulfilled.  The same is
true for the commutation relations of the level 1 generators with
the level 3 generators, which also involve generically the level 4
generators unless the indices take only at most 8 distinct values
(which forces in particular the level-3 generators to have one
repetition, i.e., to correspond to real roots).

{}Finally, the level 5 generators and the level-6 generators, which
occur in the commutation relations of level 2 with level 3, and
level 3 with itself,  involve also representations associated with
Young tableaux having a column with 9 or 10 boxes
\cite{NicFisch,Fisch}. For these to be absent, the indices must
again take on at most 8 distinct values. The commutation relations
reduce then to those of $SO(16)$, known to be valid.

\subsection{`Spin 1/2' representation}
We note that if one keeps in the above generators (\ref{k-ab}),
(\ref{k-abc}), (\ref{k-abcdef}) and (\ref{level3}) only the terms
in which the index $m$ does not transform, one gets the `spin 1/2'
representation investigated in \cite{deBuylHP}. A notable feature
of that representation is that it does not see the level-3
generators associated with imaginary roots, as one sees from
(\ref{null}).

It should be stressed that up to level 3, the commutation
relations of the $K(E_n)$ subgroups are all very similar for $n
\geq 8$ (\cite{West2,Keur3,NicCompact}). A more complete analysis
of the `spin 3/2' and `spin 1/2' representations of $K(E_{9})$
will be given in \cite{Paulot}.

\section{Extended $E_{8}$ Invariance of Supergravity Lagrangian}
\label{lagrangian}

The fermionic part of 11-dimensional supergravity is \be e^{(11)}(-\frac{1}{2}
\bpsi_\mu \gamma^{\mu \rho \nu} D_\rho \psi_\nu- \frac{1}{96}\bpsi_\mu \gamma^{\mu \nu \alpha \beta \gamma \delta }
\psi_\nu F_{\alpha \beta \gamma \delta}- \frac{1}{8} \bpsi^\a \gamma^{\gamma\delta} \psi^\b F_{\a\b \gamma \delta}
)\,, \label{sugra11}\ee where $e^{(11)}$ is the determinant of the
spacetime vielbein  and where we have dropped the terms with four
fermions. We want to compare this expression with the Lagrangian
(\ref{covariant}), where the $K(E_{10})$ representation is the spin
3/2 one constructed in the previous section.  If we expand the
Lagrangian (\ref{covariant}) keeping only terms up to level 3 and
using the dictionary of \cite{DHN} for the $K(E_{10})$ connection,
we get (see Eq. (8.7) of \cite{deBuylHP}) \bea -\frac{i}{2} \psi^T_m
\gamma^{mn} (\dot{\psi}_n - \frac{1}{2}\omega_{ab}^{R} k^{ab}.\psi_n
- \frac{1}{3!}F_{0abc} k^{abc}.\psi_n -\frac{e}{4! \,
6!}\varepsilon_{abcdp_1p_2 \cdots p_6}
F^{abcd}k_{p_1p_2 \cdots p_6}.\psi_n \nn \\
- \frac{e}{2.2!\, 8!} C^a{}_{rs} \epsilon^{rsbcdefghi}
k_{a;bcdefghi}.\psi_n) \label{sigmaf}\eea where $M$ at this level is
given by $\gamma^{mn}$ and where $\omega_{ab}^{R} = -{1 \over
2}(e_a{}^\m \dot{e}_{\m b} - e_b{}^\m \dot{e}_{\m a}) $.  In
(\ref{sigmaf}), $e$ is the determinant of the spatial vielbein,
$e^{(11)} = N \, e$ with $N$ the lapse.

We have explicitly checked the matching between (\ref{sigmaf}) and
(\ref{sugra11}).  In order to make the comparison, we
\begin{itemize}
\item  take the standard lapse $N$ equal to $e$; \item split the
eleven dimensional supergravity Lagrangian (\ref{sugra11}) into
space and time using a zero shift ($N^k = 0$) and taking the
so-called time gauge for the vielbeins $e^a_\m$, namely no mixed
space-time component; \item rescale the fermions $\psi_n
\rightarrow e^{1/2} \psi_n$ as in the spin 1/2 case, so that
$\psi_n$ in (\ref{sigmaf}) is $e^{1/2} \psi_n$ in (\ref{sugra11});
\item take the gauge choice $\psi'_0=0$ (\ref{redef});
\item take the spatial gradient of the fermionic
fields equal to zero (these gradients would appear at higher
levels); \item assume that the spatial metric is (at that order)
spatially homogeneous (i.e., neglect its spatial gradients in the
adapted frames) and that the structure constants $C^a_{\; \;bc} = -
C^a_{\; \; cb}$ of the homogeneity group are traceless (to match the
level 3 representation),
$$C^a_{\; \; ac}=0.$$
\end{itemize}
We have also verified that the matrix $M$ is indeed invariant up
to that level.

As for the spin 1/2 case, the matching between (\ref{sigmaf}) and
(\ref{sugra11}) fully covers level $3$ under the above condition of
tracelessness of $C^a_{\; \;bc}$, including the imaginary roots. For
the spin 1/2 case, this is rather direct since the null root part
vanishes, but this part does not vanish for the spin 3/2. However,
the dictionary of \cite{DHN} is reliable only for extended $E_8$.

{}Finally, we recall that the covariant derivative of the
supersymmetry spin 1/2 parameter is also identical with the
$K(E_{10})$ covariant derivative up to level 3 \cite{deBuylHP}, so
that the supersymmetry transformations are $K(E_{10})$ covariant.
The $K(E_{10})$ covariance of the supersymmetry transformations
might prove important for understanding the $K(E_{10})$ covariance
of the diffeomorphisms, not addressed previously.  Information on
the diffeomorphisms would follow from the fact that the graded
commutator of supersymmetries yields diffeormorphisms
(alternatively, the supersymmetry constraints are the square roots
of the diffeomorphisms constraints \cite{Bunster}).

\section{Conclusions}
In this paper, we have shown that the gravitino field of
11-dimensional supergravity is compatible with the conjectured
hidden $E_{10}$ symmetry up to the same level as in the bosonic
sector. More precisely, we have shown that the fermionic part of the
supergravity Lagrangian take the form (\ref{covariant}) with the
correct $K(E_{10})$ covariant derivatives as long as one considers
only the connection terms associated with the roots in extended
$E_8$, for which the dictionary relating the bosonic supergravity
variables to the sigma-model variables has been established. The
computations are to some extent simpler than for the bosonic sector
because they involve no dualization.  In sigma-model terms, the
supergravity action is given by the (first terms of the) action for
a spinning particle on the symmetric space $E_{10}/K(E_{10})$, with
the internal degrees of freedom in the `spin 3/2' representation of
$K(E_{10})$ (modulo the 4-fermion terms).

This action takes the same form as the action for a Dirac spinor
with the appropriate Pauli couplings that make it $K(E_{10})$
covariant \cite{deBuylHP}, where this time the internal degrees of
freedom are in the `spin 1/2' representation. We can thus analyse
its dynamics in terms of the conserved $K(E_{10})$ currents along
the same lines as in \cite{deBuylHP} and conclude that the BKL limit
holds.

Although the work in this paper is a necessary first step for
checking the conjectured $E_{10}$ symmetry, much work remains to be
done to fully achieve this goal.  To some extent, the analysis
remains a bit frustrating because no really new light is shed on the
meaning of the higher levels.  Most of the computations are
controlled by $E_8$ and manifest $sl_{10}$ covariance. In
particular, the imaginary roots, which go beyond $E_8$ and height
29,  still evade a precise dictionary. The works in \cite{BG} and in
\cite{DN2005} are to our knowledge the only ones where imaginary
roots are discussed and are thus particularly precious and important
in this perspective.

We have treated explicitly the case of maximal supergravity in this
paper, but a similar analysis applies to the other supergravities,
described also by infinite-dimensional Kac-Moody algebras (sometimes
in non-split forms \cite{HJ,Fre}).

\subsection*{Aknowledgments}
SdB would like to thank Sandrine Cnockaert for useful discussions.
This work is partially supported by IISN - Belgium (convention
4.4505.86), by the ``Interuniversity Attraction Poles Programme --
Belgian Science Policy'' and by the European Commission FP6
programme MRTN-CT-2004-005104, in which we are associated to
V.U.Brussel.

After this work was completed, we received the interesting preprint
\cite{DKN} where the same problem is analyzed.

\end{document}